# Tailoring magnetism of multifunctional Mn$_x$Ga films with giant perpendicular anisotropy

L. J. Zhu, D. Pan, S. H. Nie, J. Lu, J. H. Zhao[a)]

*State Key Laboratory of Superlattices and Microstructures, Institute of Semiconductors, Chinese Academy of Sciences, P. O. Box 912, Beijing 100083, China*

Abstract: We report wide-range composition and annealing effects on magnetic properties of Mn$_x$Ga films grown on GaAs (001) by molecular-beam epitaxy. We obtained single-crystalline Mn$_x$Ga films in a surprisingly wide composition range from $x$=0.76 to 2.6. We show that the magnetism could be effectively tailored by adjusting composition and annealing. Especially, when 0.76≤$x$≤1.75, Mn$_x$Ga films simultaneously show magnetization from 130 to 450 emu/cc, perpendicular anisotropy among 8.6 to 21 Merg/cc, intrinsic coercivity from 4.38 to 20.1 kOe, normal coercivity up to 3.6 kOe, energy product up to 3.4 MGOe and thermal-stability up to at least 350 °C in contact with GaAs.

[a)]Author to whom correspondence should be addressed;
Electronic mail: jhzhao@red.semi.ac.cn

Magnetic materials simultaneously with high perpendicular anisotropy ($K_u$), coercivity ($H_c$) and energy products (($BH$)$_{max}$) have great application potential in ultrahigh-density perpendicular magnetic recording, permanent magnets and spintronic devices including spin-transfer-torque (STT) magnetic random access memory (MRAM) and oscillators, and magnetoresistance sensors.[1-7] On the other hand, magnetic materials compatible with semiconductors in epitaxy allow for direct integration of magnetic devices with electronics or optical underlying circuitry.[8] Moreover, ferromagnetic metal/semiconductor heterostructures have great potential in novel applications such as spin-based light emitting diodes and field effect transistors.[8-11] Thus, it is desirable to develop novel kind of perpendicular magnetized materials epitaxied on semiconductors and further tailor their magnetic properties to meet corresponding demands for various functional applications.

In the past two decades, noble-metal-free and rare-earth-free Mn$_x$Ga films with $L1_0$ ($x$<2) or $D0_{22}$ (2<$x$<3) structures (Fig. 1(a)) have attracted increasing attention for their theory-predicted magnetic properties remarkably desirable for applications in ultrahigh-density magnetic recording, high-performance spintronic devices and economical permanent magnets. $L1_0$-MnGa ($D0_{22}$-Mn$_3$Ga) alloys with (001)-orientation are theoretically expected to have $K_u$ of 26 (20) Merg/cc, magnetization ($M_s$) of 845 (305) emu/cc, ($BH$)$_{max}$ of 28.2 (3.67) MGOe, spin polarization of 71% (88%) at the Fermi level and Gilbert damping constant of 0.0003 (0.001), respectively.[12-15] To realize practical applications of this kind of material in magnetic devices compatible with semiconductor circuitry, it is essential to explore the growth of single-crystalline $L1_0$ and $D0_{22}$ Mn$_x$Ga films with (001)-orientation on semiconductors. So far, many attempts have been made on the growth of Mn$_x$Ga films on insulators (e.g. MgO and Al$_2$O$_3$) and metals (e.g. Cr and Pt).[15-19] However, there is few study about $D0_{22}$-Mn$_x$Ga films on semiconductors. Meanwhile, although the growth of $L1_0$-Mn$_x$Ga has been studied on many semiconductors including GaN, ScN, GaSb, Si and GaAs, only a few of those films on ScN ($x$=1.5) and GaAs ($x$=1.2-1.5) demonstrated perpendicular easy axis with small magnetization and coercivity well below 230 emu/cc and 8.8 kOe at room temperatures, respectively.[8,10,18,20-22] Also, there has been a shortage of the detailed perpendicular anisotropy of these $L1_0$-MnGa films. We recently presented $L1_0$-Mn$_{1.5}$Ga films on GaAs (001) with $M_s$, $H_c$, $K_u$ and ($BH$)$_{max}$ up to 270.5 emu/cc, 42.8 kOe, 21.7 Merg/cc and 2.6 MGOe, respectively.[2] However, $M_s$ and ($BH$)$_{max}$ of these films were still smaller than theory-predicted values.[12-15] It is therefore highly desirable to investigate some effective ways to control and optimize the magnetism for various potential applications, such as by tuning composition and post-growth annealing.

There have been some reports on the composition-dependence of magnetic properties of Mn$_x$Ga polycrystalline bulks,[14,23] in which it is difficult to estimate the magnetic anisotropy and intrinsic magnetism. Also, all growth of Mn$_x$Ga films ever performed was limited to very narrow composition range belonging to either $L1_0$ or $D0_{22}$ phases.[8,10,15-18,20-22] Recently, (001)-orientated Mn$_x$Ga with $x$=1.2-3 and (111)-orientated Mn$_x$Ga with $x$=0.96-2.03 have been studied on Cr-MgO and GaN, respectively.[19,22] However, there have not been systematic studies on wide-range composition effects on magnetic properties of Mn$_x$Ga single-crystalline films with (001)-orientation, especially those on semiconductors. Moreover, an effective tailoring of magnetic properties by post-growth annealing could be expected since they are very sensitive to growth temperature.[2] Still, there are few detailed studies devoted to post-growth annealing effects on magnetic properties in Mn$_x$Ga films.

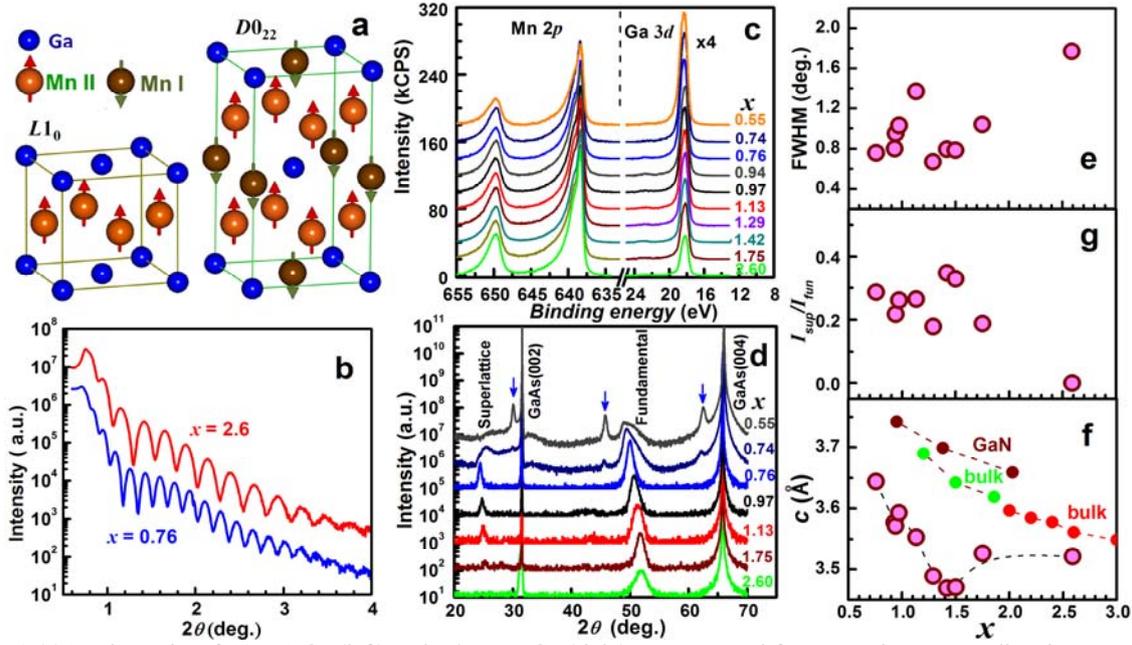

FIG. 1 (a) Lattice units of $L1_0$-MnGa (left) and $D0_{22}$-Mn$_3$Ga (right), arrows stand for magnetic moment directions. Typical (b) XRR curves, (c) High-sensitivity XPS, (d) Synchrotron XRD patterns, (e) FWHM of Mn$_x$Ga fundamental peaks, (f) $I_{sup}/I_{fun}$ and (g) $c$ of Mn$_x$Ga films. The blue arrows in (d) point to the (002), (003) and (004) peaks of MnGa$_4$ phase appeared in the Mn$_{0.55}$Ga and Mn$_{0.74}$Ga. The $c$ value for Mn$_x$Ga bulks in ref. 14 (red ●) and ref. 23 (green ●) and films grown on GaN in ref. 21 (wine ●) are also plotted in (g), for comparison with films grown on GaAs (001) (wine O).

In this letter, we present synthesis of (001)-orientated single-crystalline Mn$_x$Ga films with giant perpendicular anisotropy epitaxied on GaAs (001) in a surprisingly wide composition from $x$=0.76 to 2.6, and effective tailoring of the magnetic properties by controlling composition and annealing. These results would be much helpful for understanding this kind of materials and approaching functional applications in magnetic recording, spintronic devices and permanent magnets.

A series of Mn$_x$Ga films with different Mn/Ga atom ratio $x$ were deposited at 250 °C on 150 nm-GaAs-buffered semi-insulating GaAs (001) substrates by molecular-beam epitaxy under base pressure less than 1×10$^{-9}$ mbar.[2] After cooling down to room temperature, each film was capped with a 1.5 nm-MgO layer to prevent oxidation. The composition was designed by carefully controlling the Mn and Ga fluxes during growth and later verified by high-sensitivity x-ray photoelectron spectroscopy (XPS) measurements (Thermo Scientific ESCALAB 250Xi) with Al $K\alpha$ source and relative atomic sensitivity factors of 13.91 (1.085) for Mn 2$p$ (Ga 3$d$). The thicknesses and crystalline structures of the films were respectively determined by synchrotron radiation x-ray reflection (XRR) and x-ray diffraction (XRD) with wave length of 1.54 Å. The magnetization measurements were performed by superconducting quantum interference devices (SQUID) up to 5 T and physical properties measurement system (PPMS) up to 14 T. The annealing studies were performed at temperatures from 100 to 500 °C under vacuum below 1×10$^{-8}$ mbar.

Figure 1(b) shows typical XRR curves with the strong oscillations, suggesting sharp interfaces and good homogeneity of these samples. From XRR curves, the thicknesses were also determined to range from 35 to 50 nm. Figure 1(c) shows examples of high-sensitivity XPS of Mn 2$p$ and Ga 3$d$ in the Mn$_x$Ga films. From XPS results, $x$ was estimated to be 0.55, 0.74, 0.76, 0.93, 0.94, 0.97, 1.07, 1.13, 1.29, 1.42, 1.75 and 2.60, respectively. The asymmetry of these peaks should be attributed to the significant shielding effect to core levels by the high state-density at the Fermi level due to the alloy nature of these films. Furthermore, no considerable peak shifts are observable for both the Mn 2$p$ and Ga 3$d$ spectrums in all these films because of the protection of MgO cap layers from oxidation.

Figure 1(d) shows examples of XRD $\theta$-2$\theta$ patterns of Mn$_x$Ga films with different $x$. For $x \geqslant 0.76$, only sharp (001) ((002)) superlattice peaks and (002) ((004)) fundamental peaks of $L1_0$ ($D0_{22}$) Mn$_x$Ga films can be observed in the range from 20° to 70° besides the peaks of GaAs substrates, indicating that these are all (001)-textured single-crystalline films. For film with $x$=0.74, three small peaks irrelative to $L1_0$ (or $D0_{22}$) appear, which become dominating for film with $x$= 0.55. The new phase could be best fitted by cubic MnGa$_4$[24] with enlarged $c$ axis of 3.91 Å, while Mn$_2$Ga$_5$, MnGa$_6$, Mn$_3$Ga$_5$ and Mn$_5$Ga$_7$ phase[24,25] are not the case here. Noticeably, the (001)-orientated $D0_{22}$-Mn$_{2.6}$Ga film is of very poor crystalline quality and chemical ordering, indicating that it is difficult to crystallize the metastable $D0_{22}$-Mn$_x$Ga with high quality on GaAs, even when the lattice mismatch is very small (~2.2%). The full width at half maximum (FWHM) of fundamental

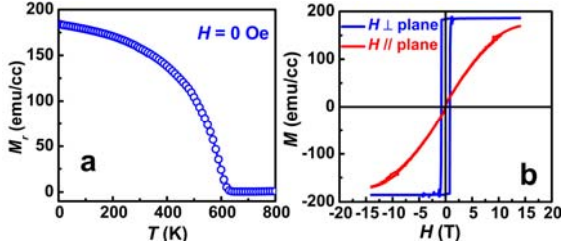

FIG. 2 (a) Temperature dependence of remanent magnetization ($M_r$) and (b) Hysteresis loops of $Mn_{1.4}Ga$ film measured at 300 K.

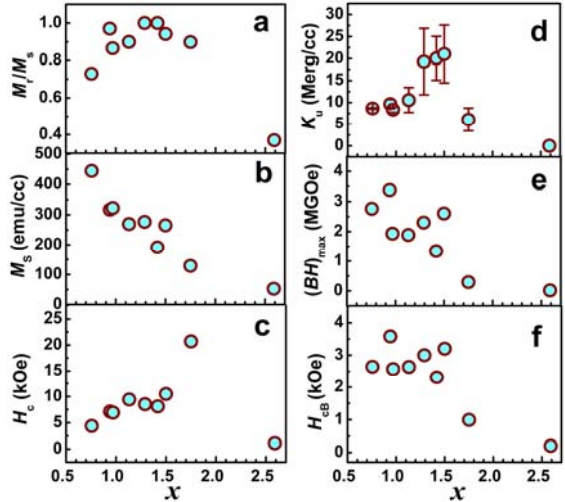

FIG. 3 (a) $M_r/M_s$, (b) $M_s$, (c) $H_c$, (d) $K_u$, (f) $H_{cB}$ and (g) $(BH)_{max}$ of $Mn_xGa$ films on GaAs (001) plotted as a function of $x$.

peaks, the integrated intensity ratio of superlattice and fundamental peaks $I_{sup}/I_{fun}$ (proportional to ordering parameters) and out-of-plane lattice parameters $c$ ($c/2$) of $L1_0$ ($D0_{22}$) $Mn_xGa$ determined from the XRD results are summarized as a function of $x$ in Figs. 1(c), (d) and (e), respectively. With exception of $Mn_{2.6}Ga$ which is of poor quality and ordering, $Mn_xGa$ films with $x \geq 0.76$ show comparable FWHM and $I_{sup}/I_{fun}$, implying similar crystalline quality and long range ordering degree. These films exhibit much short $c$-axis than bulks with same compositions probably due to tensile strains from epitaxy with GaAs. $c$ first drops dramatically, consistent with the trend in bulks[14,23] and film on GaN[21]; while it slowly comes back toward bulk values when $x$ further increases. This non-monotonic trend probably relates to both tensile strains from epitaxy on GaAs and complicated composition-dependence of high level of dislocation induced by strain relaxation (see *Supplemental Information*).

Figure 2(a) shows an example of temperature dependence of perpendicular remanent magnetization ($M_r$) of $Mn_xGa$ film deposited on GaAs (001) with Curie temperature of 630 K. In contrast, the in-plane $M_r$-$T$ curves shows nearly zero $M_r$ values in the whole temperature range (not shown here). The similar feature holds for all the films with $x$=0.76-2.60, revealing that all these films over the large composition range are magnetically phase-pure with perpendicular easy axis. Figure 2(b) shows typically both perpendicular and in-plane hysteresis loops of $Mn_xGa$ films. The perpendicular $M$-$H$ curves show square-like shapes, whereas in-plane curves exhibit almost anhysteretic loops and high saturation field even exceeding 14 Tesla. The strong difference between the perpendicular curves and in-plane curves reveals giant perpendicular anisotropy in these films. In consistence with XRD results, $Mn_xGa$ films with $x$=0.55 and 0.74 exhibit weak perpendicular anisotropy with very close $M_r$-$T$ and $M$-$H$ curves in perpendicular and in-plane directions due to the coexisting of $MnGa_4$ phase and $L1_0$-phase.

Figures 3(a)-(c) display the $x$-dependent magnetic properties including squareness ($M_r/M_s$), $M_s$ and $H_c$ determined from the 300 K perpendicular $M$-$H$ curves of these 250 °C-grown films. The $Mn_xGa$ films with $x$=0.97-1.75 exhibit high $M_r/M_s$ exceeding 0.90; but those with $x$ deviated from this range show dramatically decreased $M_r/M_s$, the tendency of which is just opposite to that of $c$ in Fig. 1(g), indicating that $M_r/M_s$ in this kind of material could be strongly affected by strains. As shown in Fig. 3(b), $M_s$ decreases dramatically from 450 to 52 emu/cc with $x$ increases from 0.76 to 2.60, which should be attributed partly to the increase of antiferromagnetic coupling between Mn atoms at different sites,[12] partly to the increase of strains induced by short $c$ axis.[2,22] The maximum $M_s$ of 445 emu/cc in $Mn_{0.76}Ga$, is still below the calculated value of 845 emu/cc for stoichiometric $L1_0$-MnGa[17] and experimental value of 600 emu/cc in 400-450 °C-annealed $Mn_{1.17}Ga$ grown on MgO [19] probably due to the low growth temperature. $H_c$ climbs up from 4.38 to 20.1 kOe as $x$ increases from 0.76 to 1.75, and then drops to 1.1 kOe at $x$=2.60. This change tendency is consistent well with that of $K_u$ estimated in Fig. 3(d) following the method used in ref. 2. $K_u$ exhibits large values from 8.6 to 21.0 Merg/cc in $L1_0$ range, while quickly degrades to 0.02 Merg/cc in $D0_{22}$-$Mn_{2.6}Ga$ because of increased disorder.

As ever discussed,[2,14] rare-earth-free and noble-metal-free $Mn_xGa$ alloys are considerable for economical permanent magnet application. Since intrinsic coercivity $H_c$ determined from $M$-$H$ curves, normal coercivity $H_{cB}$ determined from $B$-$H$ curves and magnetic energy products $(BH)_{max}$ are three key features of quality of permanent magnets, we further calculated $H_{cB}$ and $(BH)_{max}$ in Fig. 3(e) and (f). With increasing $x$ in the phase-pure range, both $H_{cB}$ and $(BH)_{max}$ decrease nearly monotonically, corresponding to the decrease of $M_s$. The film with $x$ = 1.07 exhibit the highest $H_{cB}$ and $(BH)_{max}$ of 3.6 kOe and 3.4 MGOe, respectively. The maximum of 3.4 MGOe is larger than that recent reported in $Mn_{1.5}Ga$ film (2.6 MGOe)[2] and ferrite magnets (3 MGOe),[3] which makes $L1_0$-$Mn_xGa$ alloys with low Mn compositions still promising to be developed for cost-effective and high-performance permanent magnets applications.

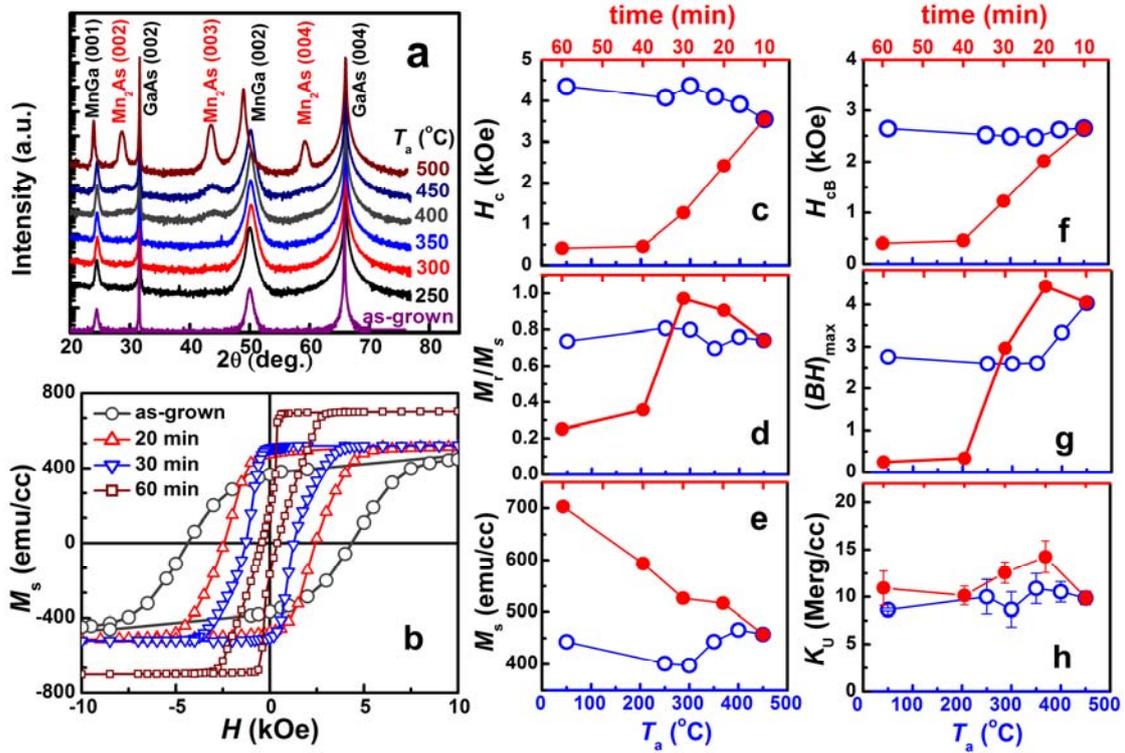

FIG. 4 (a) XRD of 10 min-annealed $Mn_{0.76}Ga$ films at different temperature. (b) Perpendicular hysteresis loops of $Mn_{0.76}Ga$ films annealed at 450 °C for different time. Annealing temperature $T_a$ (blue O, 10 min) and time (red ●, 450 °C) dependence of (c) $M_r/M_s$, (d) $M_s$, (e) $H_c$, (f) $H_{cB}$, (g) $(BH)_{max}$ and (h) $K_u$ of $Mn_{0.76}Ga$ films.

Taking into consideration of all the magnetic properties in Fig. 3, low-Mn-composition $L1_0$ $Mn_xGa$ films are more advantageous than those $D0_{22}$ film to be developed for applications in magnetic recording with areal density over 10 Tb/inch², high-performance nanoscale spintronic devices like MRAM bits with size down to several nm and economical permanent magnet applications, since they simultaneously exhibit high $M_r/M_s$, $M_s$, $K_u$, $H_c$, $H_{cB}$ and $(BH)_{max}$.

In order to investigate the thermal-dynamical properties and further tailor the magnetic properties of $Mn_xGa$ films, especially the Ga-rich single-crystalline films, we performed post-growth annealing experiments taking $L1_0$-$Mn_{0.76}Ga$ film as an example, which also exhibits the largest $M_s$ among all the as-grown films with different $x$ (Fig. 2). Figure 4(a) shows XRD patterns of as-grown and 10 min-annealed $Mn_{0.76}Ga$ films, from which we found the $Mn_{0.76}Ga$ film itself is thermal-dynamically stable up to 450 °C, and could keep stable up to at least 350 °C in contact with GaAs. This result roughly agrees with previous report that $Mn_xGa$ with $x<1.5$ should keep stable on GaAs below 400 °C.[26] Interestingly, very weak $Mn_2As$ (003) peak as a result of interfacial reaction is observable by high-sensitivity synchrotron XRD measurement, which may be not detectable by common XRD with Cu Kα radiation.[26] As summarized in Figs. 4(c)-(h), all the magnetic properties only exhibit very weak dependence on $T_a$ in the case of 10-min annealing even when a small mount of $Mn_2As$ formed at the interface after 450 °C-annealing. Fortunately, the stable temperature of at least 350 °C in respect to both structure and magnetic properties could still satisfy the demand of integration with metal-oxide-semiconductor transistors.

It should be motioned that further annealing at 450 °C for a longer time would significantly affect the magnetic behaviors of $Mn_{0.67}Ga$ films as show in Fig. 4(b). We should attribute this change mainly to both the structural improvement of $Mn_2Ga$ layer and increasing $Mn_2As$ formed at the interface with substrates. As summarized in Figs. 4 (c)-(h), with annealing time increase from 10 to 60 min, $H_c$ drops monotonically from 4 to 1 kOe; $M_s$ increases from 445 to 700 emu/cc; the highest $M_r/M_s \approx 1$ and highest $K_u$ were obtained after annealed at 450 °C for 20-30 min. Therefore, for typical $L1_0$-$Mn_{0.76}Ga$ film, annealing at 450 °C for 20-30 min would be helpful to applications like magnetic recording and STT devices which favor high $M_r/M_s$, $M_s$, $K_u$ and moderate $H_c$ at the same time. From the viewpoint of permanent magnet applications, the best should be annealing at 450 °C for 10-20 min since the largest $H_{cB}$ and $(BH)_{max}$ of 4.4 MGOe were obtained in this range.

In summary, both the wide-range composition and detailed annealing effects on the structure and magnetic properties of perpendicularly magnetized $Mn_xGa$ epitaxial films on GaAs (001) have been systematically investigated. We show that single-crystalline $Mn_xGa$ films with $L1_0$ or $D0_{22}$ ordering could be crystallized at 250 °C on GaAs along (001) direction in a surprisingly wide composition range from $x=0.76$ to 2.6. Noticeably, $L1_0$-ordered $Mn_xGa$ films showed high $M_s$, $M_r/M_s$, $K_u$, $H_c$, $H_{cB}$ and $(BH)_{max}$, which make this kind of materials favorable for applications in ultrahigh-density perpendicular magnetic recording, permanent magnets, high-performance spintronic devices. In contrast, $D0_{22}$-$Mn_{2.6}Ga$ films exhibited lower perpendicular anisotropy and weaker magnetism due to poor quality and high disordering. Annealing studies revealed thermal-stability of $Mn_xGa$ up to at least 350 ºC in contact with GaAs and effective tailoring of magnetic properties by controlling the formation of $Mn_2As$ at the interface through prolonged annealing at 450 ºC. These results would be much helpful for understanding this kind of material and approaching the functional applications in magnetic recording, spintronic devices and permanent magnets.

We gratefully thank G. Q. Pan at the U7B beamline of National Synchrotron Radiation Laboratory (NSRL) and W. Wen at the U7B beamline of Shanghai Synchrotron Radiation Facility (SSRF) in China for their help with XRD and XRR measurements. This work was supported by the NSFC under Grant Nos. 60836002 and 10920101071.

**References**
[1] C. Chappert, A. Fert, and F. N. V. Dau, Nature Mater. **6**, 813 (2007).
[2] L. J. Zhu, S. H. Nie, K. K. Meng, D. Pan, J. H. Zhao, and H. Z. Zheng, Adv. Mater. **24**, 4547 (2012).
[3] O. Gutfleisch, M. A. Willard, E. Brück, C. H. Chen, S. G. Sankar, and J. P. Liu, Adv. Mater. **23**, 821 (2011).
[4] S. Ikeda, K. Miura, H. Yamamoto, K. Mizunuma, H. Gan, M. Endo, S. Kanai, J. Hayakawa, F. Matsukura, and H. Ohno, Nature Mater. **9**, 721(2010).
[5] D. Houssameddine, U. Ebels, B. Delaët, B. Rodmacq, I. Firastrau, F. Ponthenier, M. Brunet, C. Thirion, J. P. Michel, L. Prejbeanu-Buda, M. C. Cyrille, O. Redon, and B. Dieny, Nature Mater. **6**, 447 (2007).
[6] F. B. Mancoff, J. H. Dunn, B. M. Clemens, and R. L. White, Appl. Phys. Lett. **77**, 1879 (2000).
[7] T. J. Nummy, S. P Bennett, T. Cardinal, and D. Heiman, Appl. Phys. Lett. **99**, 252506 (2011).
[8] M. Tanaka, J. P. Harbison, J. DeBoeck, T. Sands, B. Philips, T. L. Cheeks, and V. G. Keramidas, Appl. Phys. Lett. **62**, 1565 (1993).
[9] C. Adelmann, J. L. Hilton, B. D. Schultz, S. McKernan, C. J. Palmstrøm, X. Lou, H. S. Chiang, and P. A. Crowell, Appl. Phys. Lett. **89**, 112511 (2006).
[10] E Lu, D. C. Ingram, A. R. Smith, J. W. Knepper, and F.Y. Yang, Phys. Rev. Lett. **97**, 146101 (2006).
[11] H. C. Koo, J. H. Kwon, J. Eom, J. Chang, S. H. Han, and M. Johnson, Science **325**, 1515 (2009).
[12] A. Sakuma, J. Magn. Magn. Mater. **187**, 105 (1998).
[13] Z. Yang, J. Li, D. Wang, K. Zhang, and X. Xie, J. Magn. Magn. Mater. **182**, 369 (1998).
[14] J. Winterlik, B. Balke, G. H. Fecher, C. Felser, M. C. M. Alves, F.Bernardi, and J. Morais, Phys. Rev. B **77**, 054406 (2008).
[15] S. Mizukami, F. Wu, A. Sakuma, J. Walowski, D. Watanabe, T. Kubota, X. Zhang, H. Naganuma, M. Oogane, Y. Ando, and T. Miyazaki, Phys. Rev. Lett. **106**, 117201 (2011).
[16] F. Wu, S. Mizukami, D. Watanable, H. Naganuma, M. Oogane, and Y. Ando, Appl. Phys. Lett. **94**, 122503 (2009).
[17] H. Kurt, K. Rode, M. Venkatesan, P. Stamenov, and J. M. D. Coey, Phys. Status. Solidi B **248**, 2338 (2011); Phys. Rev. B **83**, 020405(R) (2011).
[18] W. Feng, D. V. Thiet, D. D. Dung, Y.Shin, and S. Chob, J. Appl. Phys. **108**, 113903 (2010).
[19] S. Mizukami, T. Kubota, F. Wu, X. Zhang, T. Miyazaki, H. Naganuma, M. Oogane, A. Sakuma, and Y. Ando, Phys. Rev. B **85**, 014416 (2012).
[20] K. M. Krishnan, Appl. Phys. Lett. **61**, 2365 (1992).
[21] A. Bedoya-Pinto, C. Zube, J. Malindretos, A. Urban, and A. Rizzi, Phys. Rev. B **84**, 104424 (2011).
[22] K. Wang, A. Chinchore, W. Lin, D. C. Ingram, A. R. Smith, A. J. Hauser, and F. Yang, J. Cry. Growth **311**, 2265 (2009).
[23] T. A. Bither and W. H. Cloud, J. Appl. Phys. **36**, 1501 (1965).
[24] H.G. Meissner and K. Schubert, Z. Metallkd. **56**, 523 (1965).
[25] J. S. Wu and K. H. Kuo, Metall. Mater. Trans. A **28A**, 729 (1997).
[26] J. L. Hilton, B. D. Schultz, S. McKernan, and C. J. Palmstrøm, Appl. Phys. Lett. 84, 3145 (2004).